\documentclass[conference]{IEEEtran}
\IEEEoverridecommandlockouts
\usepackage{cite}
\usepackage{amsmath,amssymb,amsfonts}
\usepackage{algorithmic}
\usepackage{threeparttable}
\usepackage{graphicx}
\usepackage{textcomp}
\usepackage{xcolor}
\def\BibTeX{{\rm B\kern-.05em{\sc i\kern-.025em b}\kern-.08em
    T\kern-.1667em\lower.7ex\hbox{E}\kern-.125emX}}
\begin{document}

\title{Possibility of Sleep Induction using Auditory Stimulation based on Mental States
\thanks{This work was partly supported by the Institute for Information $\&$ Communications Technology Planning $\&$ Evaluation (IITP) grant funded by the Korea government (MSIT) (No. 2017-0-00451, Development of BCI based Brain and Cognitive Computing Technology for Recognizing User’s Intentions using Deep Learning; No. 2019-0-00079, Artificial Intelligence Graduate School Program, Korea University).}
}

\author{\IEEEauthorblockN{Young-Seok Kweon}
\IEEEauthorblockA{\textit{Dept. Brain and Cognitive Engineering} \\
\textit{Korea University}\\
Seoul, Republic of Korea \\
youngseokkweon@korea.ac.kr}\\

\and

\IEEEauthorblockN{Gi-Hwan Shin}
\IEEEauthorblockA{\textit{Dept. Brain and Cognitive Engineering} \\
\textit{Korea University}\\
Seoul, Republic of Korea \\
gh\_shin@korea.ac.kr}

}

\maketitle

\begin{abstract}
Sleep has a significant role to maintain our health. However, people have struggled with sleep induction because of noise, emotion, and complicated thoughts. We hypothesized that there was more effective auditory stimulation to induce sleep based on their mental states. We investigated five auditory stimulation: sham, repetitive beep, binaural beat, white noise, and rainy sounds. The Pittsburgh sleep quality index was performed to divide subjects into good and poor sleep groups. To verify the subject's mental states between initiation of sessions, a psychomotor vigilance task and Stanford sleepiness scale (SSS) were performed before auditory stimulation. After auditory stimulation, we asked subjects to report their sleep experience during auditory stimulation. We also calculated alpha dominant duration that was the period that represents the wake period during stimulation. We showed that there were no differences in reaction time and SSS between sessions. It indicated sleep experience is not related to the timeline. The good sleep group fell asleep more frequently than the poor sleep group when they hear white noise and rainy sounds. Moreover, when subjects failed to fall asleep during sham, most subjects fell asleep during rainy sound (Cohen's kappa: -0.588). These results help people to select suitable auditory stimulation to induce sleep based on their mental states.   
\end{abstract}

\begin{IEEEkeywords}
sleep, sleep induction, repetitive beep, binaural beat, white noise, rainy sound
\end{IEEEkeywords}

\section{Introduction}
Sleep is an essential process to maintain our life and health although we lose consciousness \cite{yeom2017spatio,lee2017network,lee2019connectivity}. Sleep deprivation causes a reduction of cognitive ability, depression, and impairment of motor function \cite{sdeffects,kweon2020prediction}. In addition, a recent study suggested that it might be related to diabetes and obesity \cite{sdeffects2}. Although many people know sleep is important, they have still struggled with sleep induction. Therefore, attention to sleep induction has increased for not only sleep disorder patients but also healthy people.  

To induce sleep, people have attempted various methods. One of those methods that anyone can easily use is auditory stimulation. The binaural beat (BB) is an oscillatory stimulus that is delivered at two adjacent frequencies to each ear at the same time \cite{bb}. This induces oscillation of frequency difference to brain \cite{bb2}. 6 Hz BB combined with natural sounds showed the possibility of sleep induction \cite{bbefects}. White noise (WN) is combining sounds of all different frequencies together. When WN is given, a more number of neonates fell asleep within five minutes than is not given \cite{wnbaby}. For improving sleep experience, WN is recommended to use in intensive care unit and coronary care \cite{wn1,wn2}. Natural sounds are popular to induce sleep on the internet. Rainy sound (RS) reported about 15 million views on YouTube. Monotonous tasks, including driving a car and watching an empty computer screen, induce a micro-sleep \cite{mseffect,kweon2021automatic}. We hypothesized that repetitive beep sound (RB) induces sleep like monotonous tasks. However, sleep induction effects of auditory stimulation are unclear and require more evidence in different conditions.

In this study, we investigated the sleep induction effects of five auditory stimulation: sham, RB, BB, WN, and RS. We assessed the placebo effects using sham sound which has no sound. To confirm that there is no difference in the subject's mental states between initiation of sessions, a psychomotor vigilance task (PVT) and Stanford sleepiness scale (SSS) were performed before auditory stimuli were given. After auditory stimulation, we asked subjects to report their sleep experience during auditory stimulation. We also calculated alpha dominant duration using an electroencephalogram (EEG). It was the period that alpha power was dominant during auditory stimulation. Our results suggested that there was suitable auditory stimulation based on subjects' mental states.    

\section{Materials and Methods}

\subsection{Participants}

Thirteen healthy participants (mean age: 26.69 $\pm$ 2.46;5 female) without any neurological and auditory disease took part in the experiment. Questionnaires before experiments indicated that participants did not take any medication at the time of the experimental session. The Pittsburgh Sleep Quality Index (PSQI) was performed to evaluate sleep quality \cite{psqi}. If PSQI was under 5, the subject was in the good sleep group. The poor sleep group was the subjects who showed more than PSQI of 5 \cite{psqi2}. This study was approved by the Korea University Institutional Board Review (KUIRB-2021-0155-03) and written informed consent was obtained from participants. Because of malfunction of electroencephalogram recording, one subject was excluded to analyze data. 

\begin{figure}[tbp]
\centering
\includegraphics[width=.95\linewidth]{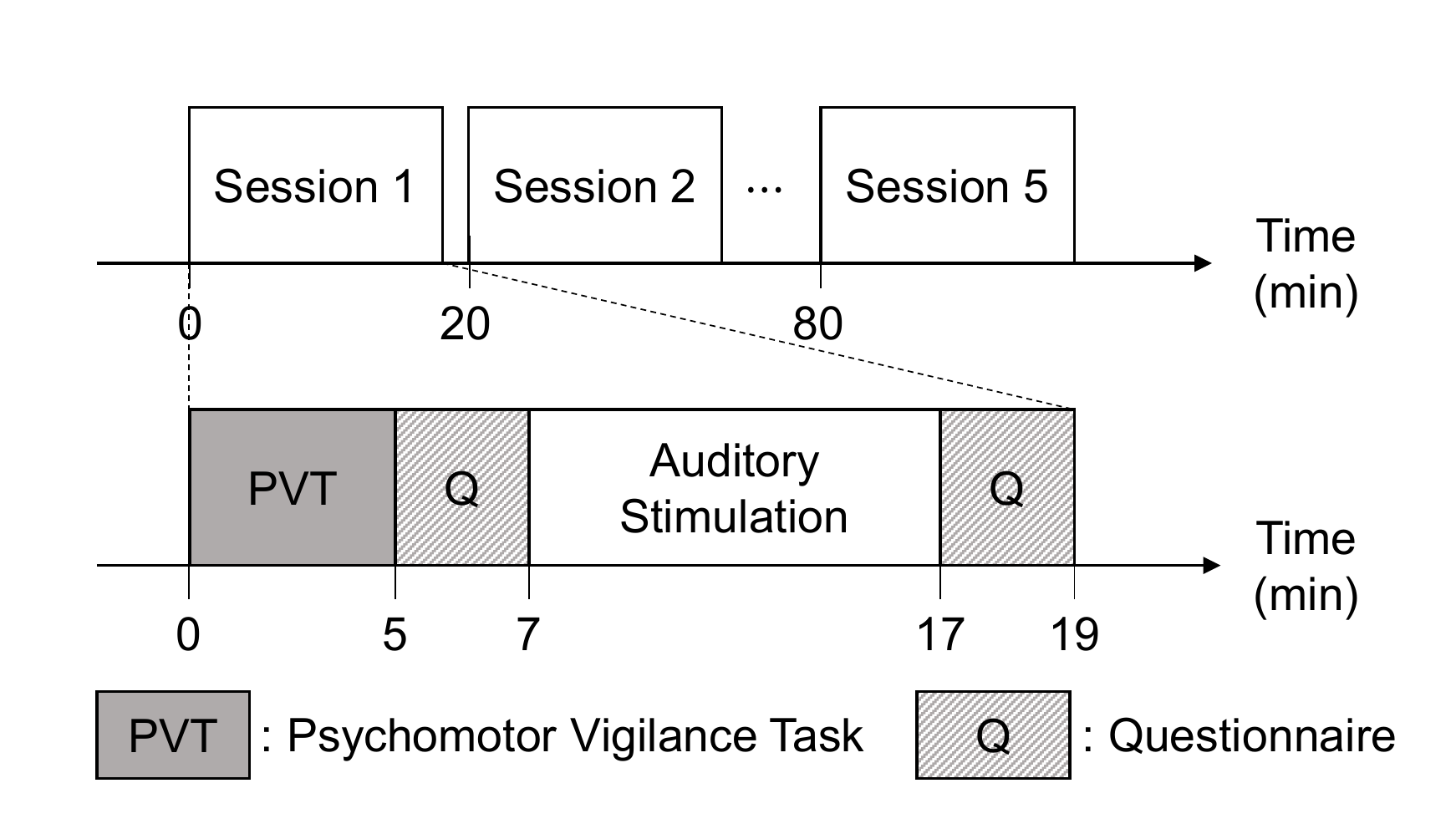}
\caption{Overview of Experiment Paradigm.}
\label{paradigm}
\end{figure}

\subsection{Stimulation and Procedures}

We prepared five auditory stimulation: sham, RB, BB, WN, and RS. During sham auditory stimulation, the audio was mute and there were not any sounds. RB was 512 Hz sound and maintained for 2 s. It was given to subjects every 5 s. The Gnaural software generated 250 Hz and 256 Hz sounds at left and right ear, respectively for BB \cite{bbefects}. We utilized free WN that was "pure noise 3" from MC2Method (https://mc2method.org/white-noise/). RS was "Rain Heavy Loud" from YouTube Studio. All subjects selected comfortable sound volume to hear auditory stimulation from 40 dB to 45 dB.

We performed modified PVT for 5 min to assess the vigilance of the participants before the auditory stimulation was given \cite{pvt_five}. At the initiative of the trial, there was nothing on the computer screen. Every 2–10 s computer screen showed the counter that represented the elapsed time from the counter was shown. Participants were instructed to stop the counter as fast as possible by pressing the space bar. After the response of participants, the elapsed time remained for 0.5 s as feedback about their reaction time.

The experiment consisted of five sessions. Before and after auditory stimulation, the SSS and Brunel mood scale ratings were done to assess participants' sleepiness and emotion. We asked subjects to report the sleep experience after auditory stimulation (Fig. \ref{paradigm}). It was reported to be 0 if subjects fell asleep, and 1 if subjects were awake. Each session randomly gave one prepared auditory stimulation. During the experiment, we always recorded the EEG of subjects with 64-channels Ag/AgCI electrode EEG setup according to the 10-20 international system using BrainAmp (ActiCap, Brain Products, Germany) \cite{lee2018high,jeong2020brain,kwon2019subject}.

\subsection{EEG Analysis}

To assess the sleep duration during auditory stimulation, we performed a spectrogram in MATLAB (R2021a, The MathWorks, Natick, MA). We set that the window is 512 and the number of overlap samples is 256. We used the four frequency bands: delta (0.5-4 Hz), theta (4-8 Hz), alpha (8-14 Hz), and beta (14-40 Hz) bands. Therefore, we got $f$ by $T$ matrix from each 10 min occipital EEG during auditory stimulation ($f$: 4 and $T$: 586). The alpha dominant duration was the ratio between the number of points that alpha was the biggest and $T$. The occipital region contained 6 channels: Oz, O1, O2, POz, PO3, and PO4.

\subsection{Statistical Analysis}

We performed one-way variance analysis to compare reaction time from PVT and SSS between sessions \cite{lee2020frontal}. Also, we used permutation test with 100 number of permutations to compare the sleep induction effects between the good and poor sleep groups. To investigate complementary usage of auditory stimulation, we calculated the Cohen's kappa value of sleep experience. If the Cohen's kappa value was close to -1, subjects experienced the opposite between auditory stimulation. Correlation between SSS and alpha dominant duration was assessed using the Kendall's tau coefficient. The significance value $\alpha$ was 0.05.

\section{Results}

\begin{figure}[tbp]
\centering
\includegraphics[width=.95\linewidth]{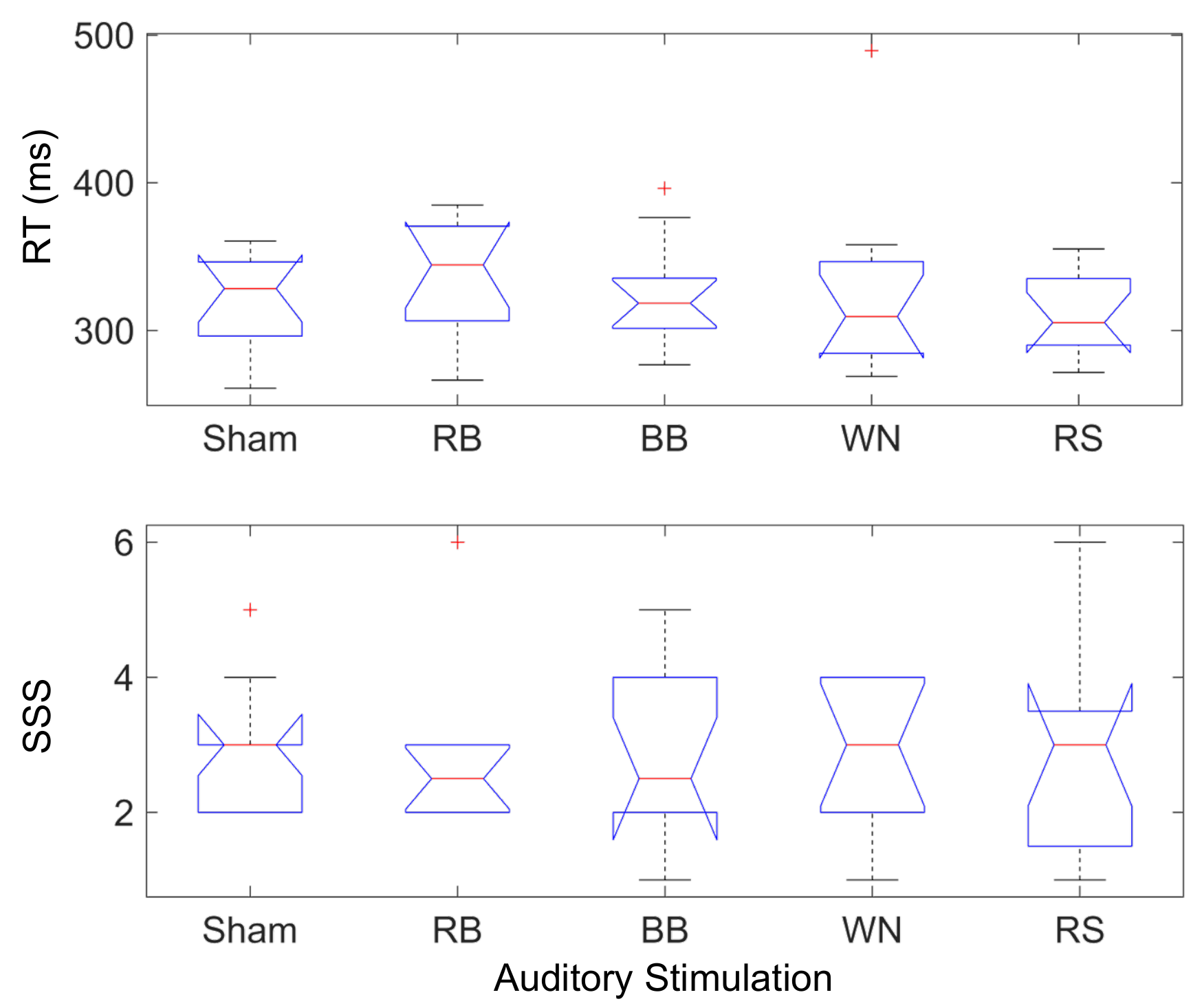}
\caption{Reaction time (RT) and Stanford sleepiness scale (SSS) before sham, repetitive beep (RB), binuaral beat (BB), white noise (WN), and rainy sound (RS) were given.}
\label{stats}
\end{figure}

\subsection{Mental States Before Auditory Stimulation}

The cognitive ability and sleepiness of subjects were not affected by time. The reaction time of PVT between sessions was not significantly different during the experiment ($p=0.855$). We also showed differences in reaction time between different auditory stimulation were not significant ($p=0.712$). Moreover, the sleepiness of subjects before auditory stimulation was not significantly different ($p=0.688$). Differences in sleepiness after auditory stimulation were not significant ($p=0.931$, Fig. \ref{stats}). 

\begin{figure}[tbp]
\centering
\includegraphics[width=.95\linewidth]{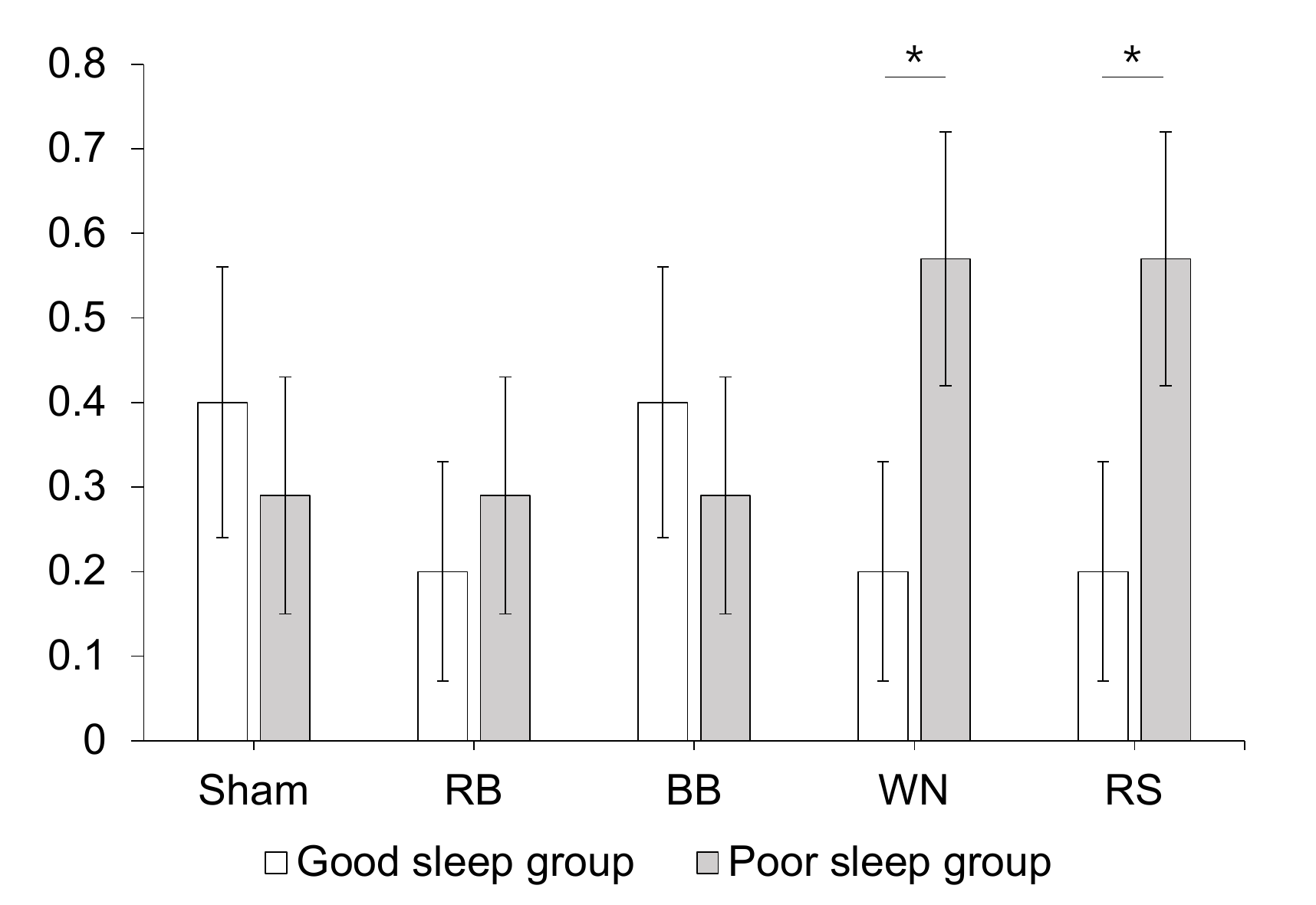}
\caption{Sleep experience of the good and poor sleep group during auditory stimulation: sham, repetitive beep (RB), binuaral beat (BB), white noise (WN), and rainy sound (RS).}
\label{sleep}
\end{figure}

\subsection{Comparison of Sleep Induction Between Sleep Groups}

Since subjects reported 0 if they fell asleep, we averaged the sleep experience of subjects. Therefore, low value meant that more people fell asleep during auditory stimulation and indicated low sleep induction effect. As you can see in the Fig. \ref{sleep}, WN and RS showed significantly low sleep induction effect to poor sleep group (WN: $p = 0.049$, RS: $p = 0.040$). However, there were no significant differences in sleep induction effects between other auditory stimulation. The good sleep group showed similar sleep induction effects to the poor sleep group during sham and BB (sham: $p = 0.604$, BB: $p = 0.475$). The poor sleep group also showed similar sleep induction effects to the good sleep group during RB ($p = 0.525$).

\begin{table}[bp]
\centering
\caption{Kappa value of sleep experience between auditory stimulation}
\label{tab:cohen}
\begin{threeparttable}
\begin{tabular}{p{0.1\linewidth}*{5}{p{0.12\linewidth}}}
\hline
     & Sham & RB & BB     & WN & RS \\ \hline
Sham & -                   & 0.400                  & 0.625                      & 0.118                  & -0.588                 \\
RB   & 0.400                    & -                  & 0.800                      & 0.636                  & -0.091                 \\
BB   & 0.625                    & 0.800                  & -                      & 0.471                  & -0.235                 \\
WN   & 0.118                    & 0.636                  & 0.471                      & -                  & -0.029                 \\
RS   & -0.588                   & -0.091                 & -0.235 & -0.029                 & -                  \\ \hline
\end{tabular}
 \begin{tablenotes}
\item[] RB, BB, WN, RS are the repetitive beeep, binuaral beat, unkown white noise, and rainy sounds, respectively.
\end{tablenotes}
\end{threeparttable}
\end{table}

\subsection{Complementary Relation Between Auditory Stimulation}

We investigated alternative auditory stimulation to induce sleep if one is failed. Table \ref{tab:cohen} showed the Cohen's kappa value of sleep experience during auditory stimulation. If subjects failed to fall asleep during sham, most subjects fell asleep during RS (kappa: -0.588). However, other auditory stimulation has no complementary relations. Sham has similar sleep induction effects to RB and BB (sham-BB kappa: 0.625, sham-RB kappa: 0.400). 

\begin{figure}[tbp]
\centering
\includegraphics[width=.95\linewidth]{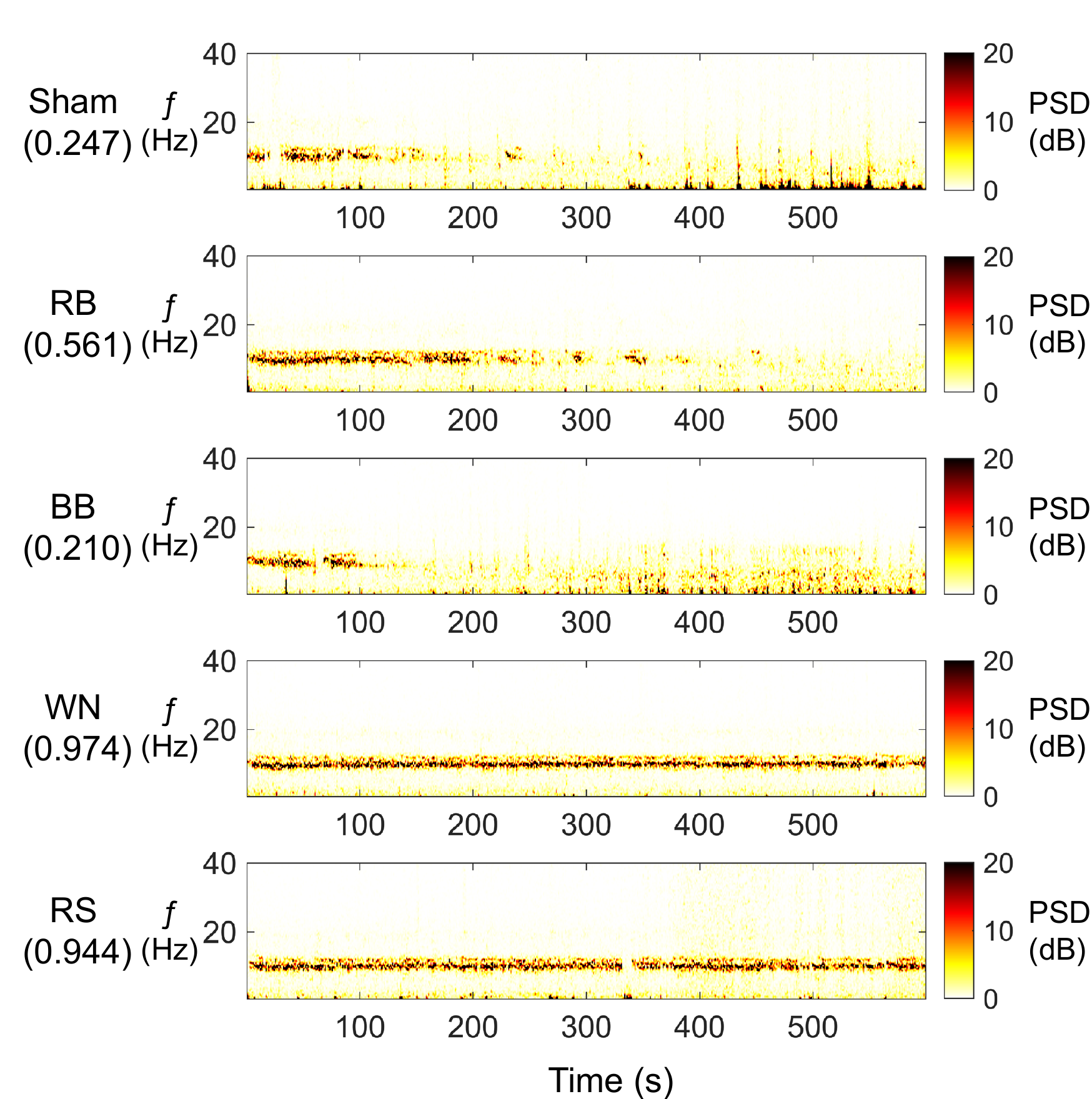}
\caption{Spectrogram of EEG in subject 8 during auditory stimulation. Alpha dominant duration was shown below type of auditory stimulation: sham, repetitive beep (RB), binuaral beat (BB), white noise (WN), rainy sound (RS).}
\label{spectro}
\end{figure}

\begin{table}[bp]
\centering
\caption{Correlation between SSS and Alpha Dominant Duration}
\label{tab:cc}
\begin{threeparttable}
\begin{tabular}{p{0.12\linewidth}*{5}{p{0.12\linewidth}}}
\hline
          & Sham                                                       & RB                                                                  & BB                                                         & WN                                                         & RS                                                         \\ \hline
\begin{tabular}[l]{@{}c@{}}Before AS\end{tabular}  & \begin{tabular}[l]{@{}c@{}}-0.017\\ (1.000)\end{tabular} & \textbf{\begin{tabular}[l]{@{}c@{}}-0.532\\ (0.036)\end{tabular}} & \begin{tabular}[l]{@{}c@{}}-0.116\\ (0.670\end{tabular}  & \begin{tabular}[l]{@{}c@{}}-0.183\\ (0.478)\end{tabular} & \begin{tabular}[l]{@{}c@{}}-0.123\\ (0.660)\end{tabular} \\
\begin{tabular}[l]{@{}c@{}}After AS\end{tabular}  & \begin{tabular}[l]{@{}c@{}}-0.307\\ (0.219)\end{tabular} & \textbf{\begin{tabular}[l]{@{}c@{}}-0.510\\ (0.040)\end{tabular}}  & \begin{tabular}[l]{@{}c@{}}-0.147\\ (0.573)\end{tabular} & \begin{tabular}[l]{@{}c@{}}-0.408\\ (0.092)\end{tabular} & \begin{tabular}[l]{@{}c@{}}-0.052\\ (0.885)\end{tabular} \\ \hline

\end{tabular}
 \begin{tablenotes}
\item[] The results are the Kendall's correlation coefficient.
\item[] Bold represents the significant correlation.
\item[] RB, BB, WN, RS are the repetitive beeep, binuaral beat, unkown white noise, and rainy sounds, respectively.
\end{tablenotes}
\end{threeparttable}

\end{table}

\subsection{Correlation Between Alpha Dominant Duration and SSS}

Alpha dominant duration was smaller when subject reported they fell asleep than they were awake during auditory stimulation in subject 8 (Fig. \ref{spectro}). Table. \ref{tab:cc} represented the correlation coefficient between SSS and alpha dominant duration. RS showed negative correlation between alpha dominant duration and SSS before and after auditory stimulation (before: $\rho = -0.532$, $p = 0.036$; after: $\rho = -0.510$, $p = 0.040$). However, there were no significant correlations between alpha dominant duration and SSS from sham, BB, WN, and RS. Sham showed lower correlations between alpha dominant duration and SSS after auditory stimulation than before (before: $-0.017$, after: $-0.307$). 

\section{Discussion}

In this study, we investigated the sleep induction effects of five auditory stimulation based on mental states. The poor sleep group showed higher sleep induction effects from WN and RS than the good sleep group. Moreover, alpha dominant duration increased if subjects reported high SSS. Another interesting finding was that there was complimentary relation between auditory stimulation. It indicated that alternative auditory stimulation could be used to induce sleep if one was failed.

Since our experimental paradigm subsequently performed sessions, you might wonder that one session's auditory stimulation could affect the subject's sleepiness and sleep deprivation to the next session. However, sleep deprivation effects and sleepiness were unchanged between sessions according to our results. Our results showed that there were no differences in reaction time and SSS between sessions. If the subjects had sleep deprivation, the reaction time of PVT increased \cite{tsdpvt}.   

RS and WN showed lower sleep induction effects to the poor sleep group than the good sleep group. There was a piece of evidence that autonomous sensory meridian response like RS was effective to induce sleep \cite{bbefects}. Their results might come from the possibility that their subjects were in the good sleep group. However, we could not check it since they did not perform PSQI. WN was recommended to use in intensive care unit and coronary care because WN has a masking effect to prevent other noise \cite{wn1,wn2}. In our laboratory environment, there was no other noise, and our results did not come from noise masking effects. However, most patients in the intensive care unit were in the poor sleep group \cite{icupsqi}. Therefore, more study needed to investigate the reason why WN was useful in the intensive care unit and coronary care.

Our results will help people to induce sleep effectively. Especially, RS will help someone who could not fall asleep without any auditory stimulation. In the future, we should investigate the two different groups: RS effective group and sham effective group. This will help people to effectively induce sleep without trial and error. The monotonous task was effective in inducing micro sleep \cite{mseffect}. RB was a monotonous task without any cognitive or motor execution. Alpha dominant duration and SSS showed a negative correlation during RB. It indicated that more sleepy subjects fell asleep longer time during RB. There might be more effective interval to present beep sounds, and we should investigate more condition. 

There were limitations to our study. First, we had a small sample to support our claims. Therefore, we will perform experiments with new subjects. Second, our experimental condition was far from the dairy sleep condition. We should investigate these results reproduced in bed at night. Finally, we should perform more advanced EEG analysis like functional connectivity \cite{zhang2019strength,zhang2017hybrid}

\section{Conclusion}
In conclusion, WN and RS were more effective in the poor sleep group than the good sleep group. RB showed high sleep induction effects if subjects were more sleepy. If someone could not fall asleep, RS helps him or her to induce sleep.

\bibliographystyle{IEEE}
\bibliography{reference}
\end{document}